\documentstyle[12pt]{article}

\def\la{\langle}
\def\ra{\rangle}

\def\lb{\lbrack}
\def\rb{\rbrack}

\def\wh{\widehat}
\def\wt{\widetilde}

\def\be{\begin{eqnarray}}
\def\ee{\end{eqnarray}}

\def\ker{\mbox{ker}}

\makeatletter
\def\section{\@startsection{section}{1}{\z@}{-3.5ex plus -1ex minus -.2ex}
{2.3ex plus .2ex}{\large\bf}}
\makeatother

\makeatletter
\@addtoreset{equation}{section}

\makeatother

\parskip=\medskipamount

\begin{document}

\vspace{8mm}

\begin{center}

{\Large \bf A remark on Schwarz's topological field theory}
\\

\vspace{12mm}

{\large David H. Adams}

\vspace{4mm}

Dept. of Pure Mathematics, Adelaide University, Adelaide, S.A. 5005, 
Australia. \\

\vspace{1ex}

email: dadams@maths.adelaide.edu.au

\vspace{7mm}

{\large Emil M. Prodanov}

\vspace{4mm}

School of Mathematics, Trinity College, Dublin 2, Ireland

\vspace{1ex}

email: prodanov@maths.tcd.ie

\vspace{3ex}

\end{center}

\begin{abstract}
The standard evaluation of the partition function $Z$ of Schwarz's 
topological field theory results in the Ray--Singer analytic torsion.
Here we present an alternative evaluation which results in $Z=1$.
Mathematically, this amounts to a novel perspective on analytic torsion:
it can be formally written as a ratio of volumes of spaces of differential
forms which is formally equal to 1 by Hodge duality. An analogous result
for Reidemeister combinatorial torsion is also obtained.

\end{abstract}

\section{Introduction}

Analytic torsion \cite{RS} arises in a quantum field theoretic context
as (the square of) the partition function of Schwarz's topological field
theory \cite{Sch(LMP),Sch(NPB)} (see \cite{AdSe(hepth)} for a detailed
review).
This has turned out to be an important result in topological quantum
field theory; for example it is used to evaluate the semiclassical
approximation for the Chern--Simons partition function \cite{W,FG-J-Roz-DA},
which gives a QFT-predicted formula for an asymptotic limit of the 
Witten--Reshetikhin--Turaev 3-manifold invariant \cite{RT} since this 
invariant arises as the partition function of the Chern--Simons gauge theory
on the 3-manifold \cite{W}. See also \cite{tft} for a review of Schwarz's
topological field theory in a general context, and 
\cite{Bytsenko-etc(NPB-MPL-hepth-hepth)} for some explicit results in the
case of hyperbolic 3-manifolds.

The partition function, $Z\,$, of Schwarz's topological field theory is 
a priori a formal, mathematically ill-defined quantity and its evaluation
\cite{Sch(LMP),Sch(NPB),AdSe(hepth)} is by formal manipulations which in 
the end lead to a mathematically meaningful result: $Z=\tau^{1/2}$
where $\tau$ is the analytic torsion of the background manifold.
In this paper we show (\S2) that there is an alternative formal evaluation
of the partition function which results in the trivial answer $Z=1$.
This result amounts to a novel perspective on analytic torsion: we find that
it can be formally written as a certain ratio of volumes of spaces of
differential forms which is formally equal to 1 by Hodge duality.

Reidemeister combinatorial torsion (R-torsion) \cite{M,RS} arises
as the partition function of a discrete version of Schwarz's topological
field theory \cite{DA(Rtorsion),DA(PRL)}. This is of potential interest
if one is to attempt to capture the invariants of topological QFT in a 
discrete, i.e. combinatorial, setting. In \S3 an analogue of
the above-mentioned result is derived for combinatorial torsion.

\section{Schwarz's topological field theory and analytic torsion}

We begin by recalling the evaluation of the partition function
\be
Z=\frac{1}{V}\int{\cal D\/}\omega\,e^{-S(\omega)}
\label{1.1}
\ee
of Schwarz's topological field theory \cite{Sch(LMP),Sch(NPB),AdSe(hepth)}.
Here $V$ is a normalisation factor to be specified below. The background
manifold (``spacetime'') $M$ is closed, oriented, riemannian, and has
odd dimension $n=2m+1$. For simplicity we assume $m$ is odd; then the 
following variant of Schwarz's topological field theory can be considered
\cite{AdSe(hepth)}: The field $\omega\in\Omega^m(M,E)$ is an $m$-form on
$M$ with values in some flat $O(N)$ vectorbundle $E$ over $M$. The
action functional is 
\be
S(\omega)=\int_M\omega\wedge{}d_m\omega\,.
\label{1.2}
\ee
Here $d_p:\Omega^p\to\Omega^{p+1}$ ($\Omega^p\equiv\Omega^p(M,E)$) is the
exterior derivative twisted by a flat connection on $E$ (which we suppress
in the notation), and a sum over vector indices is implied in (\ref{1.2})
\footnote{Note that (\ref{1.2}) vanishes if $m$ is even.}.
A choice of metric on $M$ determines an inner product in each $\Omega^p\,$,
given in terms of the Hodge operator $\ast$ by
\be
\la\omega\,,\omega'\ra=\int_M\omega\wedge\ast\omega'
\label{1.3}
\ee
Using this, the action (\ref{1.2}) can be written as 
$S(\omega)=\la\omega\,,\ast{}d_m\omega\ra$. Let $\ker(S)$ denote the radical
of the quadratic functional $S$ and $\ker(d_p)$ the nullspace of $d_p$.
Then $\ker(S)=\ker(d_m)\,$, and after decomposing the integration space
in (\ref{1.1}) as $\Omega^m=\ker(S)\oplus\ker(S)^{\perp}$ the partition
function can be formally evaluated to get
\be
Z=\frac{V(\ker(S))}{V}\,\det{}'((\ast{}d_m)^2)^{-1/4}
=\frac{V(\ker(S))}{V}\,\det{}'(d_m^*d_m)^{-1/4} 
\label{1.4}
\ee
(we are ignoring certain phase and scaling factors; see \cite{AdSe(PLB)}
for these). Here $V(\ker(S))$ denotes the formal volume of $\ker(S)$.
The obvious normalisation choice $V=V(\ker(S))$ does not preserve a
certain symmetry property which the partition function has when $S$ is
non-degenerate \cite{AdSe(hepth)}; therefore we do not use this but instead
proceed, following Schwarz, by introducing a resolvent for $S$. 
For simplicity we assume that the cohomology of $d$ vanishes, i.e.
$\mbox{Im}(d_p)=\ker(d_{p+1})$ for all $p$ ($\mbox{Im}(d_p)$ is the image 
of $d_p$). Then $S$ has the resolvent
\be
0\longrightarrow\Omega^0\stackrel{d_0}{\longrightarrow}
\Omega^1\stackrel{d_1}{\longrightarrow}\dots
\longrightarrow\Omega^{m-1}\stackrel{d_{m-1}}{\longrightarrow}\ker(S)
\longrightarrow0
\label{1.5}
\ee
which we use in the following to formally rewrite $V(\ker(S))$.
The orthogonal decompositions 
\be
\Omega^p=\ker(d_p)\oplus\ker(d_p)^{\perp}
\label{1.6}
\ee
give the formal relations
\be
V(\Omega^p)=V(\ker(d_p))\,V(\ker(d_p)^{\perp})\,.
\label{1.7}
\ee
The maps $d_p$ restrict to isomorphisms $d_p:\ker(d_p)^{\perp}
\stackrel{\cong}{\to}\ker(d_{p+1})\,$, giving the formal relations
\be
V(\ker(d_{p+1}))=|\det{}'(d_p)|\,V(\ker(d_p)^{\perp})\,.
\label{1.8}
\ee
Combining (\ref{1.7})--(\ref{1.8}) we get
\be
V(\ker(d_{p+1}))=\det{}'(d_p^*d_p)^{1/2}\,V(\Omega^p)\,V(\ker(d_p))^{-1}\,.
\label{1.9}
\ee
Now a simple induction argument based on (\ref{1.9}) and starting with
$V(\ker(S))=V(\ker(d_m))$ gives the formal relation
\be
V(\ker(S))=\prod_{p=0}^{m-1}\Bigl(\,\det{}'(d_p^*d_p)^{1/2}\,V(\Omega^p)
\Bigr)^{(-1)^p}\,.
\label{1.10}
\ee
A natural choice of normalisation is now 
\footnote{This choice can be motivated by the fact that, in an analogous 
finite-dimensional setting, the partition function then continues to exhibit
a certain symmetry property which it has when $S$ is non-degenerate
\cite{AdSe(hepth)}.}
\be
V=\prod_{p=0}^{m-1}\,V(\Omega^p)^{(-1)^p}\,.
\label{1.11}
\ee
Substituting (\ref{1.10})--(\ref{1.11}) in (\ref{1.4}) gives
\be
Z=\left\lb\,\prod_{p=0}^{m-1}\,\det{}'(d_p^*d_p)^{\frac{1}{2}(-1)^p}\right\rb
\det{}'(d_m^*d_m)^{-1/4}\,.
\label{1.12}
\ee
These determinants can be given well-defined meaning via zeta-regularisation
\cite{RS}, resulting in a mathematically meaningful expression for the
partition function. As a simple consequence of Hodge duality we have
$\det'{}(d_p^*d_p)=\det{}'(d_{n-p-1}^*d_{n-p-1})\,$, which allows to rewrite
(\ref{1.12}) as
\be
Z=\tau(M\,,d)^{1/2}
\label{1.13}
\ee
where
\be
\tau(M\,,d)=\prod_{p=0}^{n-1}\,\det{}'(d_p^*d_p)^{\frac{1}{2}(-1)^p}\,.
\label{1.14}
\ee
This is the Ray--Singer analytic torsion \cite{RS}; it is independent of the
metric, depending only on $M$ and $d$.
This variant of Schwarz's result is taken from \cite{AdSe(hepth)}; it has the
advantage that the resolvent (\ref{1.5}) is relatively simple.
The cases where $m$ need not be odd, and the cohomology of $d$ need not
vanish, are covered in \cite{Sch(LMP),Sch(NPB)} (see also \cite{AdSe(hepth)}
for the latter case). Everything we do in the following has a straightforward 
extension to these more general settings, but for the sake of simplicity 
and brevity we have omitted this.

We now proceed to derive a different answer for $Z$ to the one above.
Our starting point is (\ref{1.13})--(\ref{1.14}) which we consider as a 
formal expression for $Z\,$, i.e. we do not carry out the zeta regularisation
of the determinants. Instead, we use (\ref{1.8}) to formally write
\be 
\det{}'(d_p^*d_p)^{1/2}=\frac{V(\ker(d_{p+1}))}{V(\ker(d_p)^{\perp})}
\label{1.15}
\ee
Substituting this in (\ref{1.14}) and using (\ref{1.7}) we find
\footnote{This relation is obtained without any restriction on $m\,$,
i.e. for arbitrary odd $n$.}
\be
\tau(M\,,d)=\frac{V(\Omega^1)\,V(\Omega^3)\dots{}V(\Omega^n)}
{V(\Omega^0)\,V(\Omega^2)\dots{}V(\Omega^{n-1})}
\label{1.16}
\ee
Formally, the ratio of volumes on the r.h.s. equals 1 due to
\be
V(\Omega^p)=V(\Omega^{n-p})\,,
\label{1.17}
\ee
which is a formal consequence of the Hodge star operator being an orthogonal
isomorphism from $\Omega^p$ to $\Omega^{n-p}$. 
(Recall $\la\ast\omega\,,\ast\omega'\ra=\la\omega\,,\omega'\ra$ for all
$\omega\,,\omega'\in\Omega^p$.) This implies $Z=1$ due to (\ref{1.13}).

The formal relation (\ref{1.16}) shows that analytic torsion can be
considered as a ``volume ratio anomaly'': The ratio of the volumes on the
r.h.s. of (\ref{1.16}) is formally equal to 1, but when $\tau(M\,,d)$
is given well-defined meaning via zeta regularisation of (\ref{1.14})
a non-trivial value results in general.

It is also interesting to consider the case where $n$ is even:
In this case, using (\ref{1.7})--(\ref{1.8}) we get in place of (\ref{1.16})
the formal relation
\be
\frac{V(\Omega^0)\,V(\Omega^2)\dots{}V(\Omega^n)}
{V(\Omega^1)\,V(\Omega^3)\dots{}V(\Omega^{n-1})}
=\prod_{p=0}^{n-1}\,\det{}'(d_p^*d_p)^{\frac{1}{2}(-1)^p}=1
\label{1.18}
\ee
The last equality is an easy consequence of Hodge duality and continues 
to hold after the determinants are given well-defined meaning via
zeta regularisation \cite{RS}. On the other hand, the ratio of volumes on
the l.h.s. is no longer formally equal to 1 by Hodge duality.

\section{The discrete analogue}

Given a simplicial complex $K$ triangulating $M$ a discrete version
of Schwarz's topological field theory can be constructed which captures the
topological quantities of the continuum theory \cite{DA(Rtorsion),DA(PRL)}.
The discrete theory uses $\wh{K}\,$, the cell decomposition dual to $K\,$,
as well as $K$ itself. This necessitates a field doubling in the continuum
theory prior to discretisation: An additional field
$\omega'$ is introduced and the original action 
$S(\omega)=\la\omega\,,\ast{}d_m\omega\ra$ is replaced by the doubled action,
\be
\wt{S}(\omega,\omega')=
\Big\langle\left({\omega \atop \omega'}\right)\,,\,
\left({0 \atop \ast{}d_m} \; {\ast{}d_m \atop 0} \right)
\left({\omega \atop \omega'}\right)\Big\rangle 
=2\int_M\omega'\wedge{}d_m\omega\,.
\label{2.1}
\ee
This theory (known as the abelian BF theory \cite{tft})
has the same topological content as the original one; 
in particular its partition function, $\wt{Z}\,$, can be evaluated in an
analogous way to get $\wt{Z}=Z^2=\tau(M,d)$.
The discretisation prescription is now \cite{DA(Rtorsion),DA(PRL)}:
\be 
(\omega,\omega')&\to&(\alpha,\alpha')\in{}C^m(K)\times{}C^m(\wh{K})
\label{2.2} \\
\wt{S}(\omega,\omega')&\to&\wt{S}(\alpha,\alpha')=
\Big\langle\left({\alpha \atop \alpha'}\right)\,,\,
\left({0 \atop \ast^K{}d_m^K} \; {\ast^{\wh{K}}{}d_m^{\wh{K}} \atop 0} \right)
\left({\alpha \atop \alpha'}\right)\Big\rangle 
\label{2.3}
\ee
Here $C^p(K)=C^p(K,E)$ is the space of $p$-cochains on $K$ with values 
in the flat $O(N)$ vectorbundle $E$ and $d_p^K:C^p(K)\to{}C^{p+1}(K)$ is
the coboundary operator twisted by a flat connection on $E\;$, with 
$C^q(\wh{K})$ and $d_q^{\wh{K}}$ being the corresponding $\wh{K}$ objects;
$\ast^K:C^p(K)\to{}C^{n-p}(\wh{K})$ and
$\ast^{\wh{K}}:C^q(\wh{K})\to{}C^{n-q}(K)$ are the duality operators
induced by the duality between $p$-cells of $K$ and $(n\!-\!p)$-cells of
$\wh{K}$. The $p$-cells of $K$ and $\wh{K}$ determine canonical inner
products in $C^p(K)$ and $C^p(\wh{K})$ for each $p\,$, and with respect to
these $\ast^K$ and $\ast^{\wh{K}}$ are orthogonal maps. (The definitions
and background can be found in \cite{background}; see also \cite{RS}
and \cite{DA(Rtorsion)}.)
As in \S2 we are assuming that $m$ is odd and that the cohomology of the
flat connection on $E$ vanishes: $H^*(M,E)=0$.
Then the partition function of the discrete theory, denoted $\wt{Z}_K\,$,
can be evaluated by formal manipulations analogous to those in \S2
(see \cite{DA(Rtorsion),DA(PRL)}) and the resulting expression can be 
written as either 
\be
\wt{Z}_K=\tau(K,d^K)\qquad\ \mbox{or}\qquad\ \wt{Z}_K=\tau(\wh{K},d^{\wh{K}})
\label{2.4}
\ee
where 
\be
\tau(K,d^K)=\prod_{p=0}^{n-1}\,\det{}'(\partial_{p+1}^Kd_p^K)^{\frac{1}{2}
(-1)^p}
\label{2.5}
\ee
and $\tau(\wh{K},d^{\wh{K}})$ is defined analogously. 
Here $\partial_{p+1}^K$ denotes the adjoint of $d_p^K$ (it can be identified 
with the boundary operator on the $(p\!+\!1)$-chains of $K$).
The quantities
$\tau(K,d^K)$ and $\tau(\wh{K},d^{\wh{K}})$ coincide; in fact (\ref{2.5})
is the Reidemeister combinatorial torsion 
of $M$ determined by the given flat connection on $E\,$, and is the same for
all cell decompositions $K$ of $M$ \cite{M,RS}. (This is analogous to
the metric-independence of analytic torsion.) Moreover,
the analytic and combinatorial torsions coincide \cite{CM}, so the discrete
partition function in fact reproduces the continuum one:
\be 
\wt{Z}_K=\wt{Z}\,.
\label{2.6}
\ee

We now present an analogue of the formal argument which led to $Z=1$ in \S2.
Consider
\be
\tau(K,d^K)\,\tau(\wh{K},d^{\wh{K}})=
\prod_{p=0}^{n-1}\,\det{}'(\partial_{p+1}^Kd_p^K)^{\frac{1}{2}(-1)^p}
\det{}'(\partial_{p+1}^{\wh{K}}d_p^{\wh{K}})^{\frac{1}{2}(-1)^p}\,.
\label{2.7}
\ee
Using the analogues of (\ref{1.15}) and (\ref{1.7}) in the present 
setting, 
\be
\det{}'(\partial_{p+1}^Kd_p^K)^{1/2}
=\frac{V(\ker(d_{p+1}^K))}{V(\ker(d_p^K)^{\perp})}
\label{2.7.3} 
\ee
and
\be
V(C^p(K))=V(\ker(d_p^K))\,V(\ker(d_p^K)^{\perp})\,,
\label{2.7.5}
\ee
and the corresponding $\wh{K}$ relations,
we find an analogue of the formal relation (\ref{1.16}):
\be
&&\tau(K,d^K)\,\tau(\wh{K},d^{\wh{K}}) \nonumber \\ 
&=&\frac{V(C^1(K))\,V(C^3(K))\dots{}V(C^n(K))}
{V(C^0(K))\,V(C^2(K))\dots{}V(C^{n-1}(K))}\,
\frac{V(C^1(\wh{K}))\,V(C^3(\wh{K}))\dots{}V(C^n(\wh{K}))}
{V(C^0(\wh{K}))\,V(C^2(\wh{K}))\dots{}V(C^{n-1}(\wh{K}))} \nonumber \\
& &\label{2.8}
\ee
Formally, the r.h.s. equals 1 due to
\be
V(C^p(K))=V(C^{n-p}(\wh{K}))\,,
\label{2.9}
\ee
which is a formal consequence of the duality operator being an orthogonal
isomorphism from $C^p(K)$ to $C^{n-p}(\wh{K})$ (i.e. $\la\ast^K\alpha\,,
\ast^K\alpha'\ra=\la\alpha\,,\alpha'\ra$ for all $\alpha,\alpha'\in{}C^p(K)$).
This implies that, formally,
\be
\wt{Z}_K=\Bigl(\tau(K,d^K)\,\tau(\wh{K},d^{\wh{K}})\Bigr)^{1/2}=1\,.
\label{2.10}
\ee

Thus we see that combinatorial torsion can also be considered as a 
``volume ratio anomaly'' in an analogous way to analytic torsion.

Finally, in the $n$ even case it is straightforward to find a combinatorial
analogue of the formal relation (\ref{1.18}) --we leave this to the 
reader.

{\it Acknowledgements.} D.A. is supported by an ARC fellowship. E.P. is
supported by funding from FORBAIRT and Trinity College Dublin.

\end{document}